\documentclass{ws-p9-75x6-50}
\font\cc cmcsc10
\def\co{{\cal O}}
\def\MNRAS{\em Mon. Not. Royal Astron. Soc.}
\def\Nat{\em Nature}
\def\ASS{\em Astrophys. Space Science}
\def\ApJ{\em Astroph. Journal}
\def\ApJL{\em Astroph. Journal Lett.}
\def\AJ{\em Astron. Journal}
\def\NewA{\em New Astron.}
\def\JC{\em Journ. Comp. Phys.}
\def\CP{\em Computers in Phys.}
\def\ZA{\em Z. f. Astroph.}
\def\ARAA{\em Ann. Rev. Astron. Astroph.}
\def\AA{\em Astron. Astroph.}
\def\AAR{\em Astron. Astroph. Rev.}
\def\PASJ{\em Proc. Astron. Soc. Japan}
\def\Rev{\em Rev. Mod. Astroph.}
\def\CeMDA{\em Celest. Mech. Dynam. Astron.}

\begin{document}

\title{Towards the Million Body Problem on the Computer --
no news since the Three-Body-Problem?}

\author{R. Spurzem}

\address{Astronomisches Rechen-Institut, M\"onchhofstra\ss e 12-14,
D-69120 Heidelberg, Germany\\E-mail: spurzem@ari.uni-heidelberg.de}

\author{A. Kugel}

\address{Dept. for Computer Science V, Univ. of Mannheim, B6-26,
D-68131 Mannheim, Germany\\E-mail: kugel@ti.uni-mannheim.de}


\maketitle

\abstracts{Astrophysical Challenges which demand the solution of
the one million (or more) gravitating body problem are briefly discussed
for the fields of cosmology, galactic nuclei and globular star clusters.
Results from the classical three-body problem are to be combined with
methods of statistical mechanics or thermodynamics in order to provide
a reasonable model for astrophysically relevant particle numbers of
the order $10^4-10^8$. For high-precision modelling of star clusters
and galactic nuclei over the required hundreds and thousands of crossing
times specially tailored ``brute-force'' $N$-body integration methods
are used together with special purpose hardware. A recent implementation
of such code on a general purpose parallel computer is presented.
After a comparison of the relevance of different
$N$-body algorithms a new concept for a more flexible customized
special purpose computer based on a combination of GRAPE and FPGA
is proposed. It is an ideal machine for all kinds of $N$-body simulations
using neighbour schemes, as the Ahmad-Cohen direct $N$-body codes and
smoothed particle hydrodynamics (SPH) for systems including interstellar gas
dynamics.
}

\section{Introduction}
 Gravitation is the only one of the four fundamental forces in physics
 which cannot be shielded by particles of opposite charge and whose
 force at long range interactions does not have any (e.g. exponential)
 cutoff; the gravitational force is described by Newton's inverse square
 force law to the largest possible ranges in which classical mechanics
 applies. So it does not have a preferred scale and as a consequence
 gravitational forces play an important role for the dynamical
 evolution of astrophysical systems on practically all scales.
 Among the most challenging problems which have
 been treated by purely gravitational
 $N$-body simulations are structure formation in the universe, evolution
 of galactic nuclei and globular star clusters. Even if
 non-gravitative effects become important, like hydrodynamics or
 magnetohydrodynamic forces (e.g. for star formation or dissipative
 galaxy formation) some of the methods to solve the relevant dynamical
 equations represent the gas by particles interacting by a combination
 of gravitational and non-gravitative forces (smoothed particle
 hydrodynamics, {\cc SPH}, \cite{Lucy77,GingoldM77}). 
 Thus, simulations using particles to follow the dynamical evolution
 of astrophysical systems are one of the most important tools in
 computational astrophysics and have become a third independent
 experimental field of astrophysical
 research besides theory and observations. 

 This paper is organized as follows. First we present some key questions
 which are being addressed by present or
 future direct $N$-body simulations in the
 fields of globular star clusters, galactic nuclei, and cosmological models
 of structure formation (Sect. 2). Second, the main algorithms for
 astrophysical $N$-body simulations are briefly introduced and discussed
 in comparison with each other (Sect. 3).
 In Sect. 4 their implementation
 on special purpose hardware is discussed and
 a new concept for faster and more flexible hardware
 tailored to various kinds of direct $N$-body simulations 
 including gas dynamics is presented.

 \section{Some Astrophysical Key Questions}
 \subsection{Globular Star Clusters}

 There is an excellent review of the internal dynamics of globular star
 clusters \cite{MeylanH97}. Here we only want to stress some selected
 topics which are relevant to the subject of $N$-body simulations. Globular
 star clusters are nicely spherical (sometimes slightly flattened) star 
 clusters consisting of $10^5$ -- $10^6$ stars. 
 Since their escape velocity
 is small compared to the typical velocities of stellar winds and explosions
 they are practically gas-free. They are ideal stellar dynamical laboratories
 because their relevant thermal (two--body relaxation) and dynamical
 timescales are smaller than their lifetime. Globular cluster systems
 exist around many other galaxies as well \cite{ForbesBG97}.

 Due to their relative isolation, lack of observable interstellar gas and
 due to their symmetry globular clusters are well approximated by simplified
 theoretical models.
 Since the relaxation
 timescale is long compared to the dynamical time they develop
 through a sequence of dynamical (virial) equilibria.
 The fundamental
 kinetic equation in such case is
 the Fokker-Planck equation. The use of
 this equation for stellar dynamics was inspired by plasma physics
 \cite{Cohn80}. Recent
 models of that type include the effects of anisotropy (differences
 between radial and tangential velocity dispersion, which can be
 present even in spherical systems) \cite{Takahashi96,Takahashi97}.
 Another improvement includes for the first time the effect of
 rotation for those globular clusters which are slightly
 flattened \cite{EinselS99}.  Also anisotropic gaseous models
 based on a moment evaluation of the Fokker--Planck equation were
 successfully used \cite{LouisS91,Spurzem94}.

 In the
 presence of self-gravitation many concepts of thermodynamics,
 however, have to be used with care.
 So, for example, there is
 no global thermodynamic equilibrium, because a system of gravitating
 point masses can always achieve infinite amounts of binding energy just
 by moving two or more or all of the particles closer and closer together.
 At some limiting central velocity dispersion
 general relativity takes
 over and the cluster collapses by a dynamical instability
 towards a black hole
 \cite{ShapiroT85}. Before reaching
 that limit most realistic astrophysical systems, however, will reach
 the limit of physical collisions and merging of their stars. Another
 alternative is that
 strong two--body correlations (close binaries) form which subsequently stop
 the global gravitational collapse by
 superelastic scatterings with field stars \cite{BettwieserS84}. 
 If there are too many binaries, however,
 the fundamental assumption of
 using the one-particle distribution function and the Fokker-Planck equation
 breaks down. 

 The second obstacle of thermodynamic methods to treat astrophysical
 ensembles of particles is simply that particle numbers are not large
 enough. Therefore stochastic fluctuations, deviations
 from thermodynamic expectation values in individual representations of
 e.g. star clusters are much more 
 significant than in any laboratory gas;
 the amplitude of such fluctuations
 can be of a size comparable to that of the observed
 quantity. Hence it is by no means guaranteed, that a given 
 individual globular
 cluster consisting of some one million or less particles strictly
 evolves according to models derived from statistical mechanics.
 In a seminal series of
 papers \cite{GierszH94a,GierszH94b,GierszH96,GierszH97}
 the results of statistical models based on the Fokker-Planck
 approximation were compared
 with ensemble averages of a number of statistically independent
 direct $N$-body simulations \cite{GierszH94a,GierszH94b,GierszH96,GierszH97}. 
 From this and similar
 work \cite{SpurzemA96,Makino96} 
 one can conclude that in spherical isolated clusters statistical
 models in spherical symmetry with standard two--body relaxation
 work fairly well, but already in the case of a galactic tidal field
 with an enhanced mass loss by stellar escapers, severe
 problems occur in understanding the results of the direct $N$-body models
 and their relation to the results of the Fokker-Planck results 
 \cite{AarsethH98}.
 Note that such problems occur for one of the still most
 simplified globular cluster models; no rotation and no mass loss by
 stellar evolution was included, stars were considered as
 point masses and no effects of binary stellar evolution taken into account,
 no primordial binaries present, no time-dependent three-dimensional
 tidal field, and so on. Consequences from that are two--fold: first
 great care should be taken advancing Fokker-Planck and gaseous models
 to more complicated situations, second direct $N$-body models should be
 seen either as a theoretical tool to check and gauge the statistical
 models (eventually after a process of ensemble averaging) or they
 should employ a realistic particle number to directly model an
 individual, real star cluster. Besides the questions of gravothermal
 oscillations \cite{BettwieserS84,Makino96}
 in very large $N$-body systems and the scalability of
 cluster models in a galactic tidal field,
 we would like to stress here the importance to acquire information
 on the pre- and post-collapse evolution of 
 $N$-body models of rotating globular clusters, 
 for which very little is known yet, except a first series
 of Fokker-Planck type models \cite{EinselS99}.

 Hence only an
 exact, direct $N$-body integrator should be used
 for problems of globular cluster and galactic nuclei
 dynamics, which treats the two-body relaxation
 by small angle gravitative encounters of all impact parameters with
 a maximal accuracy and simultaneously, accurately
 and efficiently follows the formation and evolution
 of very close binaries, whose timescales differ by many orders
 of magnitude from the dynamical timescale of the whole cluster.
 Such requirements are fulfilled by {\cc Nbody5}
 \cite{Aarseth85} and its successors (see Sect. 3).
 As a final remark for
 this subsection, we stress that in very young stellar clusters,
 like newly forming globular clusters seen around merging galaxies
 \cite{Schweizeretal96}, the timescales to deplete the cluster from
 remaining protostellar gas by stellar winds, for star formation
 and evolution of massive stars, are comparable to the dynamical 
 time of the cluster. Since mass segregation by two--body relaxation
 can be faster than the standard relaxation by a factor of $M/m$, where
 $M$ are the most massive species, and $m$ is the average particle mass
 in the cluster, even two-body effects are not completely negligible
 at cluster formation. Modelling such situation in a context of 
 cooling and fragmentation of a gas cloud \cite{MurrayL96} including
 stellar dynamical effects would require an highly accurate $N$-body
 integrator in dynamical coupling with a gaseous component. It is interesting
 to note here, that the problem of star formation in molecular cloud
 complexes, which is another challenging problem of theoretical astrophysics,
 has been studied by a numerical model based on particles (SPH, smoothed
 particle hydrodynamics) and using special purpose computers (GRAPE,
 gravity pipe), on which we will elaborate below 
 \cite{KlessenBB98,Klessen97}.

 \subsection{Galactic Nuclei}

 Another long--standing problem of collisional stellar dynamics is
 the question of the equilibrium system and dynamical evolution of
 a cusp of stars surrounding a massive central black hole. Such 
 massive black holes are very likely to reside in the centres
 of galaxies as a fossile of earlier acticity \cite{KormendyR95}.
 Their formation as a result of collisionless dynamical
 general relativistic collapse and dissipative processes during
 galaxy formation is very likely but not yet fully understood
 \cite{QuinlanS90}. In an earlier paper \cite{FrankR76}
 the interplay between mass and energy transport
 by two--body relaxation and loss--cone accretion of stars on orbits
 with low angular momentum by the black hole was studied by
 semi-analytical scaling arguments; these results were 
 confirmed by Monte-Carlo numerical models
 later \cite{MarchantS80} 
 followed by multi--mass direct numerical solutions of the 
 1D Fokker--Planck equation for isotropic stellar cusps \cite{MurphyCD91}.
 Only recently the first self-consistent
 $N$-body models of massive black holes including a sufficient number
 of stars in their surrounding cusps were done by use of hybrid $N$-body
 algorithms \cite{Quinlan96,QuinlanH97} or a high-speed
 special purpose computer for a direct summation algorithm 
 \cite{MakinoE96,Makino97}.
 However, the latter work was occupied
 mainly with the question of dynamical friction of black hole binaries
 in a galactic nucleus after a merger event. Still the standard picture
 of \cite{FrankR76} has not yet carefully been checked by using a
 direct full $N$-body simulation. It is not certain, whether the assumption
 that two--body relaxation dominates the evolution is correct; it has
 been suggested that large angle close encounters of stars with each other and
 with the black hole compete with it, and that there may be non--standard
 relaxation processes at work \cite{RauchT96}. These are
 interesting open question to tackle with high accuracy pure particle
 $N$-body simulations. Even more important as in the case of globular clusters,
 however, are the possible effects of gas produced by stellar collisions,
 which can accumulate in the centre due to the much deeper central potential,
 and form new stars \cite{QuinlanS90,Rees97}.
 We would like to conclude this subsection with the
 final remark that this is again a physical situation where highly
 accurate direct $N$-body models, eventually dynamically coupled with
 the dynamics of a gas component are very important for future understanding
 of such objects.

\subsection{Cosmology and Structure Formation}

 In the standard paradigm of cosmological structure formation 
 primordial quantum
 fluctuations grow
 gravitationally in a universe dominated by non-dissipative dark matter. In
 the non-linear regime the distribution of masses can be estimated by
 simple theory \cite{PressSch74}, later extended to
 $N$-body models \cite{Whiteetal87,NavarroFW97}.
 On small scales gas
 physics, which (e.g. in the case of star formation) is only known approximately
 has to be included into the models 
 \cite{Steinmetz96}.
 Recently it has been shown, that softening
 of the gravitational potential, which was adopted in most of the models,
 causes spurious two--body relaxation effects \cite{SteinmetzW97}.
 Consistently \cite{Mooreetal98} find that
 the structure of cold-dark-matter (CDM) haloes significantly changes
 if models with much higher resolution in particle number are used.
 Again, we want to conclude here that high
 resolution, high-accuracy $N$-body simulations, gravitationally coupled
 with a gas component are useful to study such questions.

 \section{Numerical Methods of $N$-body integration}

 \begin{table}[htb]
 
\caption{Algorithms for $N$-body Simulations}
\begin{tabular}{llll}
\hline
Number & Name & Scaling \\
\hline
\noalign{\leftline{\underbar{\cc No particle-particle relaxation:}}}
\noalign{\medskip}
1 &  PM -- particle mesh  & $\co(N)+\co(n^3)$ \\
2.&  Fast Multipole &  $\co(N)+\co(nlm)$ \\
3.&  Self Consistent Field & $\co(N)+\co(nlm)$\\
\noalign{\medskip}
\noalign{\leftline{\underbar{\cc ``Exact'':}}}
\noalign{\medskip}
4.& {\cc Nbody1 - 4} & $\co(N^2)$ \\
5.& {\cc Nbody5 - 6} & $\co(NN_n)+\co(N^2)$ \\
6.& {\cc Kira}       &  \\
\noalign{\medskip}
\noalign{\leftline{\underbar{\cc ``Mixed'':}}}
\noalign{\medskip}
7.& {\cc Tree} & $\co(N\log N)$ \\
8.& {\cc P${}^3$M} &
    $\co(N_n^2)+\co(N)\co(nlm)$ \\
\hline
\end{tabular}
\end{table}
 In Table 1 an overview over the most commonly used present algorithms
 for direct $N$-body simulations is given. The symbols used in the
 ``Scaling'' column denote the particle number $N$, a neighbour number
 $N_n$ (compare Sect. 4), a grid resolution $n$ or the
 number of terms $nlm$ in a series evaluation of the gravitational
 potential. We want to comment only very briefly
 on each of the methods to give an overview for the reader. The
 first group has been labelled ``no particle--particle relaxation''
 because it does not use direct gravitational forces between particles.
 The gravitational potential is computed from the particle configuration
 via an intermediate step, either through a mesh in coordinate space or an
 orthogonal function series. Reviews on classical
 particle mesh (PM) techniques
 can be found in \cite{Sellwood87}.
 ``{\cc Superbox}'' is a multi--grid method in
 a classical PM scheme suitable for high resolution problems and
 relaxing the inflexibility of conventional PM methods somehow
 \cite{MadejskyB93}. 
 Fast multipole methods used as presented by Greengard 
 \cite{Greengard90,GreengardR87}
 can only efficiently be used for codes using the same 
 timestep for particles, which makes them unfeasible for astrophysical
 problems with gravitating particles developing into highly structured and/or
 inhomogeneous states. Codes using an orthogonal series expansion
 (so-called ``self consistent field'' or SCF codes) have been introduced
 to the astrophysical community mainly by \cite{HernquistO92},
 although there are earlier similar approaches \cite{Clutton72}. 

  In all cases where a highly accurate computation of the gravitational
 potential with all its graininess due to individual particles, responsible
 for various relaxation effects, is necessary, there is no way to avoid
 a direct brute--force summation algorithm, where individual
 pairwise forces are computed. Such approach goes back to the
 early 60's \cite{Aarseth63,vHoerner60}.
 Close encounters and the formation of binaries,
 whose binding energies are large compared to the thermal energy
 of the system have led to a special algorithm to treat 
 hierarchical subsystems in such codes (named {\cc Nbody3 - 6} \cite{Aarseth98})
 by a transformation to
 regularized variables \cite{MikkolaA98,Mikkola97,MikkolaA96}
 The more advanced code versions employ an individual or
 hierarchical block time step scheme and a high order
 time integrator
 \cite{Aarseth85,MakinoA92}, completed by
 an Ahmad-Cohen neighbour scheme \cite{AhmadC73}
 to reduce the number of full force computations
 (code versions {\cc Nbody5 - 6} \cite{Aarseth85,Aarseth98}).
 The algorithm is well
 parallelizable and has been implemented on general purpose
 parallel computers \cite{SpurzemB99}. 
 Last but not least the {\cc Kira} code is mentined in Table 1, which
 is a fresh approach for a high accuracy direct $N$-body simulation code
 and has been used in studies of the dissolution of star clusters with 
 stellar evolution in a tidal field \cite{PortegiesZetal98,PortegiesZT98}.
 At the end of Table 1 there are the ``mixed'' codes; one
 is the {\cc Tree}-Code \cite{BarnesH86},
 using particle--particle forces in principle; 
 it groups, however, subsets of particles in some distance from a test particle
 together, taking only their centres of masses into account (and if
 required also some multipole moments of the mass distribution). It is
 highly efficient for lumpy particle configurations, where the configuration
 has a small overall filling factor, and has been used very successfully
 for large-scale cosmological simulations
 and models of merging galaxies, partly even including a gas component
 treated by {\cc SPH} \cite{DaveDH97,MihosDH98}.
 Most {\cc Tree}-Code implementations do not require very high accuracies,
 for example an energy error of up to a few percent is generally tolerated.
 Enforcing in a {\cc Tree}--code very high accuracy as it is required
 for globular cluster models ($10^{-3}$ \%, a typical value achieved
 for direct Aarseth {\cc Nbody}-integrators) leads to a very significant
 reduction of its efficiency \cite{McMillanA93}. Finally, another
 ``mixed'' code also used especially for cosmological simulations with
 an {\cc SPH} gas component is the ${\rm P}^3{\rm M}$--code, 
 for which we refer to a
 recent paper of \cite{PearceC97}.

 \section{Hardware}
 \subsection{Introduction}

 The construction of special-purpose hardware to compute gravitational 
 forces in direct $N$-body simulations was inspired by the fact that
 the total CPU time required for one time step\
 of all particles scales as $T = \alpha N + \beta N^2 $, with
 some numerical time constants $\alpha$ and $\beta$. For large particle
 number $N$ the second part (the pairwise force calculation) 
 consumes most of the time \cite{Sugimotoetal90}. So our Japanese
 colleagues built the GRAPE hardware, whose development
 culminated in the presentation of the GRAPE-4 Teraflop computer
 \cite{Makinoetal97,MakinoT98}. 
 The latter could be so fast, that according to 
 Amdahl's law the other parts of the code (e.g. to advance the particles or
 compute neighbour related quantities)
 become the bottleneck of the simulation. This becomes even more serious
 if particle based methods are applied to solve the equations of 
 hydrodynamics. The widely used SPH (smoothed particle hydrodynamics)
 represents the fluid by an ensemble of particles each carrying mass and
 momentum (as in the $N$-body problem). Thermodynamic observables
 are determined in a local averaging process over a given set of
 neighbouring particles. Hence the scaling properties of such codes, 
 as well as the Ahmad-Cohen pure $N$-body codes using a neighbour scheme, 
 can be modelled by a
 timing formula for the CPU time required per timestep as
 $T = \alpha N + \beta N^2 + \delta N\cdot N_n$, where $\delta$ is another
 time constant and $N_n$ a typical neighbour number, for which neighbour
 forces (in case of {\cc Nbody6}) or gas dynamical forces (in case of
 {\cc SPH}) are to be calculated on a test particle. 

 However, the special purpose machines of the GRAPE series reach their
highest efficiency only for problems, which can be tackled with pure
and clean $N$-body algorithms such as {\cc Nbody4} or {\cc Kira}. For SPH or
standard $N$-body simulations using an Ahmad-Cohen neighbour scheme or a very
large number of close (so-called primordial) binaries, or even worse
for molecular dynamics simulations with potentials other than the
Coulomb potential (e.g. van der Waals) they are not the optimal
choice. One commonly used solution is to use general purpose massively
parallel machines as the CRAY T3E, for which a competitive
implementation {\cc Nbody6++} exists using MPI and SHMEM \cite{SpurzemB99}. 
While
its performance compares well with one of the single GRAPE-4 boards, as can
be seen in Fig.~1, a
larger scale GRAPE machine or the coming GRAPE-6 are still much more
efficient for the pure $N$-body case. There is still work in progress,
however, to improve the implementation on the general purpose parallel
computers.

\begin{figure}[htb]
\epsfig{file=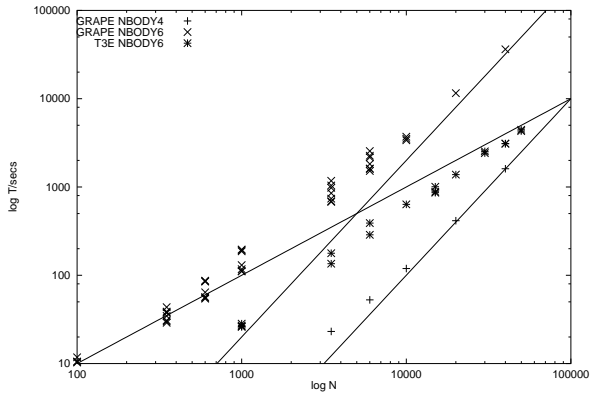,height=7.6truecm,angle=270}
\caption{
 CPU time needed for one $N$-body time unit as
 a function of particle number $N$ using the parallel version
 {\cc Nbody6++} -- on special purpose computers
 GRAPE and CRAY T3E, and {\cc Nbody4} on GRAPE (keys for symbols given in the
 top left corner of the plot).  
 For each $N$ several data points are given for measurements
 with varying average neighbour number and processor/pipeline number,
 which are not individually discriminated in the figure. Power-Laws scaling
 with $N$ and $N^2$ are given for comparison.}
 \end{figure}

The new solution presented here is to build a hybrid machine, which
uses for the intermediate range forces a reconfigurable custom
computing machine -- an FPGA processor. This new system will profit
from both the extremely high performance of the GRAPE's for the $\co(N^2)$
gravitational force computation and the high degree of flexibility of
the FPGA processor which lets it adapt to the needs of the various
hydrodynamic (SPH) oriented computations in the $\co(NN_n)$ regime.

\subsection{FPGA processors}

The family of FPGA devices was introduced in 1984 by Xilinx. FPGA's
feature a large number of relatively simple elements with configurable
interconnects and an indefinite number of reconfiguration cycles with
short configuration times. All configuration information is stored in
SRAM cells. The basic processing element (PE) of all current
mainstream FPGA's is a 4-input/1-output look-up-table (LUT) with an
optional output register. The functionality of the FPGA is thus
determined by the contents of the look-up-tables within the PE's and
the "wiring" between these elements.  Over the last few years
FPGA performance has increased tremendously as it profits from both:
Increased density by a factor of 24 from 1993 through 1998 (Xilinx
XC4000: 400 to 18400 elements) increased speed by a factor of 3 from
1994 through 1998 (Xilinx XC4000: 133 to 400MHz internal toggle rate).

8 years of experience at the University of Mannheim with FPGA based
computing machines shows that this new class of computers is an ideal
concept for constructing special-purpose processors combining both the
speed of a hardware and the flexibility of a software solution. The
so--called FPGA processors consist of a matrix of FPGA's and memory forming
the computational core. In addition there are a (programmable) I/O
unit and an internal (configurable) bus system. As processing unit,
I/O unit and bus system are implemented in separate modules, this kind
of system provides scalability in computing power as well as I/O
bandwidth.

FPGA processors have shown to provide superior performance in a broad
range of fields, like encryption, DNA sequencing, image processing,
rapid prototyping. 
The hybrid microprocessor/FPGA systems developed at the
University of Mannheim \cite{Hoegletal95} are in particular suitable for:

* acceleration of computing intensive pattern recognition tasks in
High Energy Physics (HEP) and Heavy Ion Physics,

* subsystems for high-speed and high-frequency I/O in HEP, 

* 2-dimensional industrial image processing, 

* 3-dimensional medical image visualization and 

* acceleration of
multi-particle interaction calculations in astronomy

A commonly used procedure to adjust a hybrid system to different
applications is modularity. ATLANTIS implements modularity on
different levels. First of all there are the main entities host CPU and FPGA
processor which allow to partition an application into modules
tailored for either target. Next the architecture of the FPGA
processor uses one board-type to implement mainly computing tasks and
another board-type to implement mainly I/O oriented tasks. A backplane
based interconnect system provides scalability and supports an
arbitrary mix of the two board-types. Finally modularity is used on
the sub-board level by allowing different memory types or different
I/O interfaces per board type; it is an optimal machine to be
applied to the astrophysical problems \cite{Kuberkaetal99}.

\subsection{Architecture}

Using FPGA's to accelerate complex computations using floating-point
algorithms has not been considered a promising enterprise in the past
few years. The reason is that general floating-point as well as
particular $N$-body implementation have shown only poor
performance on FPGA's. Usually $N$-body calculations and particle based
fluid simulations need a computing performance in at least Teraflop
range and are accelerated with the help of ASIC-based co-processors.
Nonetheless we have recently investigated the performance of a
certain sub-task of the SPH algorithm on the Enable++ system
\cite{Kuberka98}. The
results indicate that FPGA's can indeed provide even in this area a
significant performance increase. Computation of the kernel function and
the density determination of the SPH algorithm were implemented
on 15 out of 16 core FPGA's of the Enable++ system making
heavy use of the configurable interconnect structure.
For the implementation a 28bit floating-point format was
used: 1 sign-bit, 7 bits exponent, 20 bits mantissa. The maximum
pipeline depth is 6 stages and a result is produced at every clock
cycle. The total performance for the code in the loop is therefore
$16\cdot 13$ MHz = 208 Mflops with the XC4013-5 chips and 
$16\cdot 32$ MHz = 512 Mflops
with the XC4028-2 chip respectively. If the XC4036-3 implementation
will allow -- as we expect -- that 2 instances can run in
parallel, the performance will increase to 1.024 Mflops. Parallel I/O
is also done with 52 or 128MB/s on the input side plus a few MB/s on
the output side. ATLANTIS will support two instances of this code to
run in parallel on one computing board.

\subsection{AHA-GRAPE}

For astrophysical particle simulations including self-gravity, the
determination of the gravitational potential at each particles
position is usually the most expensive step in terms of computational
time required. This step shall be done by the special hardware GRAPE
for force computation in $N$-body simulations, which proved highly
efficient in the case of a pure point-mass simple algorithm (e.g. 
{\cc Nbody1})
case. For many more realistic applications however, some parts of the
code become important bottlenecks if the gravitational force
calculation is done very fast. They are usually of order $\co(NN_n)$ --
where $N_n$ is a neighbour particle number --- and comprise

1. Computation of the neighbour force when using SPH or a more complex,
but more efficient $N$-body algorithm (about 20 flops per pairwise
force, of which order $N_n$ per particle per time-step have to be
computed).

2. Computation of the kernel function in SPH, its derivatives, and
terms related to gas dynamical quantities (about 100 flops per
pairwise particle interaction, of which again $\co(N_n)$ per particle per
time-step have to be computed ).

3. Integration of binary motion in regularized coordinates as a
function of near perturbers (order $N_n$); this is a very sophisticated
algorithm and cannot easily be estimated now in its complexity
\cite{Mikkola97,MikkolaA98}.

4. Integration of SCF force (self-consistent field \cite{HernquistO92} 
to compute
approximately the gravitational potential of distant particles ) for
hybrid $N$-body models .

The floating point operations related to 1) and 2) are in principle
straightforward to map onto an FPGA processor, however critical for
the performance is the word length which is sufficient for each
component of the sum of pairwise forces. Test calculations have to be
performed to clarify this.  The two subsystems -- GRAPE cluster and
FPGA processor -- will be connected to the host workstation by the
PCI bus, either directly or via an interface. Within the subsystems
the respective local buses will be used to broadcast sample data and
intermediate results. In a second step and by close cooperation with
the Tokyo group a hierarchical coupling of the FPGA device with GRAPE,
including memory and control of the GRAPE, could be envisaged. This
will further improve performance by parallelization of force
computations and data communication. Fig. 2 displays the performance
estimates for various systems with and without FPGA processor.
The system with FPGA is called AHA-GRAPE 
({\bf A}daptive {\bf H}ydrodyn{\bf A}mics {\bf GRAPE}), since it is
ideally suited for high accuracy gravitational $N$-body simulations
and hydrodynamical models using the SPH algorithm. 

\begin{figure}[htb]
\epsfig{file=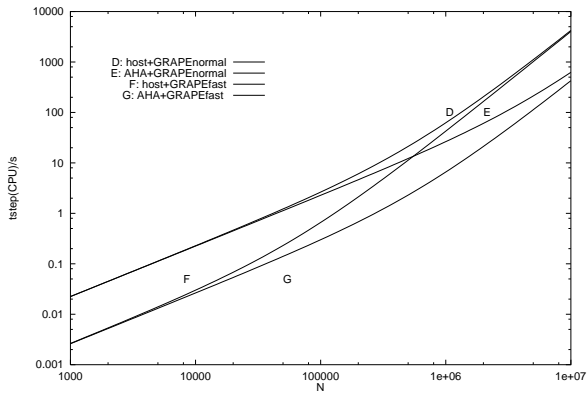,height=7.6truecm,angle=270}
\caption{CPU time per step required for a simulation with direct
 gravitational force computation and neighbour scheme
 (SPH gas dynamics or Ahmad-Cohen $N$-body code)
 as a function of particle number for the
 proposed AHA-GRAPE machine (E, G) and a standard GRAPE-host combination
 (D, F), for a ``normal'' (D, E) and ``fast'' (F, G) GRAPE. Details see
 main text.}
\end{figure}

\subsection{Status and plans}

At present (February 1999) a test implementation of
the SPH-loop/step1 on Enable++ is carried out to verify the estimated
performance. By mid 1999 the new ATLANTIS system will be available where
the full SPH-code has to be implemented. A communication library for
LINUX must be developed, supporting simultaneous transfers between
host/GRAPE and host/FPGA respectively. We expect the first prototype
AHA-GRAPE system to be available in mid 2000. The key figures for this
prototype are 50Mflops for the host workstation, 5Gflops for the FPGA
processor and 500Gflops for the GRAPE subsystem, which have
been used for the timing estimates depicted in Fig.~2.  The presence of the
FPGA processor will lead to an increase in performance by a factor of
10 and will allow us to handle up to $\approx 10^6$ particles in
collision dominated $N$-body simulations and a few $10^7$ particles in SPH.

\section*{Acknowledgments}

It is a great pleasure to acknowledge in the name
of the GRAPE user community in Germany the
help and support received from D. Sugimoto, now continued by
J. Makino
and the members of his team, which originated from the time 
Professor Sugimoto visited as
a Gauss Professor G\"ottingen observatory in 1983. 
Support by the DFG (German Science Foundation) under grants Sp 345/9-1,2
and Sonderforschungsbereich 439 (Galaxies in the Young Universe) is
gratefully acknowledged. 
Computing time at NIC J\"ulich (former HLRZ), 
H\"ochstleistungsrechenzentrum Stuttgart (HLRS) and
SSC Karlsruhe is gratefully acknowledged.

\end{document}